\documentclass[slac_one]{revtex4}
\usepackage{graphicx}
\usepackage{fancyhdr}
\renewcommand{\headrulewidth}{0pt}
\renewcommand{\footrulewidth}{0pt}

\def\plb#1#2#3{{\it Phys.\ Lett.\ }{\bf B#1}, #2 (#3)}
\def\prd#1#2#3{{\it Phys.\ Rev.} {\bf D#1}, #2 (#3)}

\newcommand{\prlb}[3]{{\it Phys.\ Rev.\ Lett.} {\bf #1}, #2 (20#3)}
\newcommand{\epj}[3]{{\it Euro.\ Phys.\ J.} {\bf C#1}, #2 (20#3)}
\newcommand{\etal}{{\it et al.}}

\newcommand{\optbar}[1]{\shortstack{{\tiny (\rule[.4ex]{1em}{.1mm})} 
  \\ [-.7ex] $#1$}}		
\def\poptm{\raise.8ex\hbox{+}%
    \kern-0.9em\lower.6ex\hbox{{\tiny (}--{\tiny )}}}	
\def\lsim{\;\raisebox{-.6ex}{$\stackrel{<}{\sim}$}\;}

\newcommand{\ra}{\rightarrow}
\newcommand{\beq}{\begin{equation}}
\newcommand{\eeq}{\end{equation}}
\newcommand{\Eq}[1]{Eq.~(\ref{eq#1})}

\begin{document}

\title{Neutrino Physics\footnote{Lectures presented at the 2004 SLAC Summer Institute. To appear in the Proceedings. Fermilab publication FERMILAB-PUB-05-236-T.}}

\author{Boris Kayser\footnote{Email address: boris@fnal.gov}}
\affiliation{Fermilab, Batavia IL 60510, USA}

\begin{abstract}
Thanks to compelling evidence that neutrinos can change flavor, we now know that they have nonzero masses, and that leptons mix. In these lectures, we explain the physics of neutrino flavor change, both in vacuum and in matter. Then, we describe what the flavor-change data have taught us about neutrinos. Finally, we consider some of the questions raised by the discovery of neutrino mass, explaining why these questions are so interesting, and how they might be answered experimentally,
\end{abstract}

\maketitle

\thispagestyle{fancy}

\section{PHYSICS OF NEUTRINO OSCILLATION}
 \label{s1}

\subsection{Introduction}\label{s1.1}

There has been a breakthrough in neutrino physics. It has been discovered that neutrinos have nonzero masses, and that leptons mix. The evidence for masses and mixing is the observation that neutrinos can change from one type, or ``flavor'', to another. In this first section of these lectures, we will explain the physics of neutrino flavor change, or ``oscillation'', as it is called. We will treat oscillation both in vacuum and in matter, and see why it implies masses and mixing.

That neutrinos have masses means that there is a spectrum of neutrino mass eigenstates $\nu_i, i=1,2,\ldots$, each with a mass $m_i$. What leptonic mixing means may be understood by considering the leptonic decays $W^+ \ra \nu_i + \overline{\ell_\alpha}$ of the $W$ boson. Here, $\alpha = e, \mu$, or $\tau$, and $\ell_e$ is the electron, $\ell_\mu$ the muon, and $\ell_\tau$ the tau. The particle $\ell_\alpha$ is referred to as the charged lepton of flavor $\alpha$. Mixing means simply that in the $W^+$ decays to the particular charged lepton $\overline{\ell_\alpha}$, the accompanying neutrino mass eigenstate is not always the {\em same} $\nu_i$, but can be {\em any} of the different $\nu_i$. The amplitude for $W^+$ decay to produce the specific combination $\overline{\ell_\alpha} + \nu_i$ is denoted by $U^*_{\alpha i}$. The neutrino state emitted in $W^+$ decay together with the particular charged lepton $\overline{\ell_\alpha}$ is then
\beq
|\nu_\alpha > = \sum_i U_{\alpha i}^* | \nu_i > ~~ .
\label{eq1}
\eeq
This superposition of mass eigenstates is called the neutrino of flavor $\alpha$.

The quantities $U_{\alpha i}$ may be collected into a matrix known as the leptonic mixing matrix \cite{r1}. According to the Standard Model, $U$ is unitary. This unitarity guarantees that when the neutrino $\nu_\alpha$ interacts in a detector and creates a charged lepton, the latter will always be $\ell_\alpha$, the charged lepton with the same flavor as the neutrino. That is, a $\nu_e$ always yields an $e$, a $\nu_\mu$ a $\mu$, and a $\nu_\tau$ a $\tau$.

The relation (\ref{eq1}), expressing a neutrino of definite flavor as a superposition of mass eigenstates, may be inverted to express each mass eigenstate $\nu_i$ as a superposition of flavors:
\beq
|\nu_i > = \sum_\alpha U_{\alpha i} | \nu_\alpha > ~~ .
\label{eq2}
\eeq
The flavor-$\alpha$ fraction of $\nu_i$ is obviously $|U_{\alpha i}|^2$. When $\nu_i$ interacts in a detector and produces a charged lepton, this flavor-$\alpha$ fraction is the probability that the charged lepton will be of flavor $\alpha$.

We turn now to the physics of neutrino oscillation.

\subsection{Neutrino Flavor Change in Vacuum}\label{s1.2}

\begin{figure}[!hbtp]
\begin{center}
\includegraphics[scale=0.8]{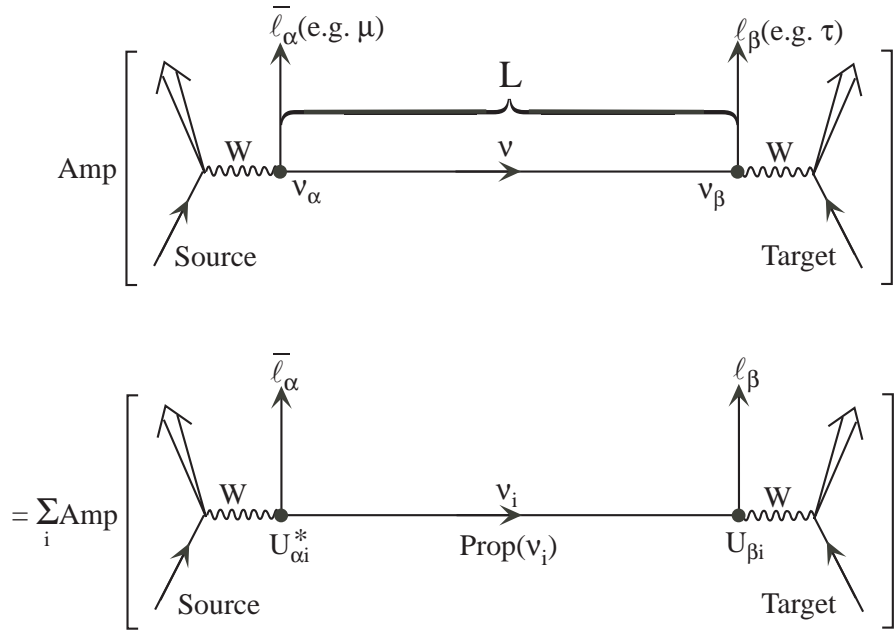}
\caption{Neutrino flavor change (oscillation) in vacuum. ``Amp'' denotes an amplitude.} 
\label{f1} 
\end{center}
\end{figure}

A typical neutrino flavor change, or ``oscillation'', is depicted schematically in the top part of Fig.~\ref{f1}.
There, a neutrino source produces a neutrino together with a charged lepton $\overline{\ell_\alpha}$ of flavor $\alpha$. Thus, at birth, the neutrino is a $\nu_\alpha$. It then travels a distance $L$ to a detector. There, it interacts with a target and produces a second charged lepton $\ell_\beta$ of flavor $\beta$. Thus, at the time of its interaction in the detector, the neutrino is a $\nu_\beta$. If $\beta \neq \alpha$ (for example, if $\ell_\alpha$ is a $\mu$ but $\ell_\beta$ is a $\tau$), then, during its journey to the detector, the neutrino has morphed from a $\nu_\alpha$ into a $\nu_\beta$.

This change of neutrino flavor, $\nu_\alpha \ra \nu_\beta$, is a quintessentially quantum-mechanical effect. Indeed, it entails some quantum-mechanical subtleties that are still debated to this day \cite{r2}. However, there is little debate about the `` bottom line''---the expression for the flavor-change probability, P($\nu_\alpha \ra \nu_\beta$). Therefore, in the interest of brevity, here we will derive this expression using an efficient approach \cite{r3} that contains all the essential quantum physics, even though it may not do justice to the subtleties.

Since, as described by \Eq{1}, a $\nu_\alpha$ is actually a coherent superposition of mass eigenstates $\nu_i$, the particle that propagates from the neutrino source to the detector in Fig.~\ref{f1} is one or another of the $\nu_i$, and we must add the contributions of the different $\nu_i$ coherently. Thus, with ``Amp'' denoting an amplitude, Amp($\nu_\alpha \ra \nu_\beta$) is given by the lower part of Fig.~\ref{f1}. There, the contribution of each $\nu_i$ is a product of three factors. The first is the amplitude for the neutrino produced together with an $\overline{\ell_\alpha}$ by the source to be, in particular, a $\nu_i$. As has been said, this amplitude is $U_{\alpha i}^*$. The second factor is the amplitude for the produced $\nu_i$ to propagate from the source to the detector. This factor is denoted by Prop($\nu_i$) in Fig.~\ref{f1}, and will be determined shortly. The final factor is the amplitude for the charged lepton created by the $\nu_i$ when it interacts in the detector to be, in particular, an $\ell_\beta$. From the Hermiticity of the Hamiltonian that describes neutrino-charged lepton-$W$ boson couplings, it follows that if Amp($W \ra \overline{\ell_\alpha} \nu_i ) = U_{\alpha i}^*$, then Amp$(\nu_i \ra \ell_\beta W) = U_{\beta i}$. Thus, the final factor in the $\nu_i$ contribution is $U_{\beta i}$, and
\beq
\mathrm{Amp}(\nu_\alpha \ra \nu_\beta) = \sum_i U_{\alpha i}^* \mathrm{Prop}(\nu_i) U_{\beta i} ~~ .
\label{eq3}
\eeq

Now, what is Prop($\nu_i$)? To find out, we go to the $\nu_i$ rest frame. We call the time in that frame $\tau_i$. If $\nu_i$ has rest mass $m_i$, then in its rest frame its state vector obeys the trivial Schr\"{o}dinger equation
\beq
i \frac{\partial}{\partial \tau_i}|\nu_i (\tau_i)>\; = m_i |\nu_i (\tau_i)> ~~ .
\label{eq4}
\eeq
The solution to this equation is obviously
\beq
|\nu_i (\tau_i)>\; = e^{-im_i\tau_i}|\nu_i (0)> ~~ .
\label{eq5}
\eeq
Thus, the amplitude for $\nu_i$ to propagate for a time $\tau_i$, which is just the amplitude $<\nu_i (0)|\nu_i (\tau_i)>$ for finding the original $\nu_i$ state $|\nu_i (0)>$ in the time evolved state $|\nu_i (\tau_i)>$, is simply $\exp [-im_i\tau_i]$. Prop($\nu_i$) is just this amplitude with $\tau_i$ the proper time taken by $\nu_i$ to travel from the neutrino source to the detector.

For Prop($\nu_i$) to be useful to us, we must re-express it in terms of laboratory-frame variables. Two of these variables are the laboratory-frame distance, $L$, that the neutrino travels between its source and the detector, and the laboratory-frame time, $t$, that elapses during the trip. The value of $L$ is chosen by the experimenters through their choices for the location of the source and the location of the detector. Similarly, the value of $t$ is chosen by the experimenters through their choices for the time when the neutrino is created and the time when it is detected. Thus, $L$ and $t$ are defined by the experiment, and are common to all $\nu_i$ components of the beam. Different $\nu_i$ do not have different values of $L$ and $t$ from each other.

The other two laboratory-frame variables are the energy $E_i$ and momentum $p_i$ of mass eigenstate $\nu_i$ in the laboratory frame. By Lorentz invariance, the phase $m_i\tau_i$ in the $\nu_i$ propagator Prop($\nu_i$) is given in terms of the laboratory-frame variables by
\beq
m_i \tau_i = E_i t - p_i L~~.
\label{eq6}
\eeq

The reader might object that, in reality, neutrino sources are essentially constant in time, and the time $t$ that elapses between the birth of a neutrino and its detection is not measured. This objection is quite correct. In practice, a realistic experiment averages over the time $t$ taken by the neutrino during its journey. Now, suppose two components of the neutrino beam, one with (laboratory-frame) energy $E_1$ and the other with energy $E_2$, contribute coherently to the neutrino signal observed in the detector. If the time taken by the neutrino to travel from source to detector is $t$, then by the time the beam component with energy $E_j \; (j=1,2)$ reaches the detector, it has picked up a phase factor $\exp [-iE_j t]$. Thus, the interference between the $E_1$ and $E_2$ components of the beam will involve a phase factor $\exp [-i(E_1 - E_2) t]$. Averaged over the unobserved travel time $t$, this factor vanishes, {\em unless $E_2 = E_1$}. Thus, the only components of a neutrino beam that contribute coherently to a neutrino oscillation signal are components that have the same energy \cite{r4}. In particular, the different mass eigenstate components of a beam that contribute coherently to the oscillation signal must have the same energy, $E$.

At energy $E$, mass eigenstate $\nu_i$, with mass $m_i$, has a momentum $p_i$ given by
\beq
p_i = \sqrt{E^2 - m_i^2} \cong E - \frac{m_i^2}{2E} ~~ .
\label{eq7}
\eeq
Here, we have used the fact that, given the extreme lightness of neutrinos, $m_i^2 \ll E^2$ for any realistic energy $E$. From Eqs.~(\ref{eq6}) and (\ref{eq7}), we see that at energy $E$ the phase $m_i \tau_i$ in Prop($\nu_i$) is given by
\beq
m_i \tau_i \cong E(t-L) + \frac{m_i^2}{2E}L ~~ .
\label{eq8}
\eeq
In this expression, the phase $E(t-L)$ is irrelevant since it is common to all the interfering mass eigenstates. Thus, we may take
\beq
\mathrm{Prop}(\nu_i) = \exp [-im_i^2 \frac{L}{2E}] ~~ .
\label{eq9}
\eeq

Using this result, it follows from \Eq{3} that the amplitude for a neutrino to change from a $\nu_\alpha$ into a $\nu_\beta$ while traveling a distance $L$ through vacuum with energy $E$ is given by
\beq
\mathrm{Amp}(\nu_\alpha \ra \nu_\beta) = \sum_i U_{\alpha i}^* \, e^{-im_i^2 \frac{L}{2E}} U_{\beta i} ~~ .
\label{eq10}
\eeq
This expression holds for any number of flavors and mass eigenstates. Squaring it, we find that the probability P($\nu_\alpha \ra \nu_\beta$) for $\nu_\alpha \ra \nu_\beta$ is given by
\begin{eqnarray}
\mathrm{P}(\nu_\alpha \ra \nu_\beta) & = & |\mathrm{Amp}(\nu_\alpha \ra \nu_\beta)|^2 \nonumber \\
	& = & \delta_{\alpha\beta} - 4\sum_{i>j} \Re (U^*_{\alpha i} U_{\beta i} U_{\alpha j}U^*_{\beta j}) \sin^2 (\Delta m^2_{ij}\frac{L}{4E}) \nonumber \\
	 & & \phantom{\delta_{\alpha\beta}} + 2\sum_{i>j} \Im (U^*_{\alpha i} U_{\beta i} U_{\alpha j}U^*_{\beta j}) \sin\, (\Delta m^2_{ij}\frac{L}{2E}) ~~ ,
\label{eq11}
\end{eqnarray}
where
\beq
\Delta m_{ij}^2 \equiv m_i^2 - m_j^2 ~~ .
\label{eq12}
\eeq
In obtaining \Eq{11}. we have made judicious use of the unitarity of $U$.

The oscillation probability P($\nu_\alpha \ra \nu_\beta$) of \Eq{11} is that for a {\em neutrino}, rather than an {\em antineutrino}, as one can see from Fig.~\ref{f1}, which shows that the oscillating particle is produced in association with a charged {\em antilepton} $\bar{\ell}$, and produces a charged {\em lepton} $\ell$ in the detector. The probability P($\overline{\nu_\alpha} \ra \overline{\nu_\beta}$) for the corresponding antineutrino oscillation may be found from P($\nu_\alpha \ra \nu_\beta$) using the fact that the process $\overline{\nu_\alpha} \ra \overline{\nu_\beta}$ is the CPT-mirror image of $\nu_\beta \ra \nu_\alpha$. Thus, assuming that CPT invariance holds,
\beq
\mathrm{P}(\overline{\nu_\alpha} \ra \overline{\nu_\beta}) = \mathrm{P}(\nu_\beta \ra \nu_\alpha) ~~ .
\label{eq13}
\eeq
Now, from \Eq{11} we see that 
\beq
\mathrm{P}(\nu_\beta \ra \nu_\alpha;\: U) = \mathrm{P}(\nu_\alpha \ra \nu_\beta; \: U^*) ~~ .
\label{eq14}
\eeq
Thus, if CPT invariance holds, it follows from \Eq{11} that
\begin{eqnarray}
\mathrm{P}(\optbar{\nu_\alpha} \ra \optbar{\nu_\beta}) & = & \delta_{\alpha\beta} - 4\sum_{i>j} \Re (U^*_{\alpha i} U_{\beta i} U_{\alpha j}U^*_{\beta j}) \sin^2 (\Delta m^2_{ij}\frac{L}{4E}) \nonumber \\
	 &  & \phantom{\delta_{\alpha\beta}} {\rm \poptm}\; 2\sum_{i>j} \Im (U^*_{\alpha i} U_{\beta i} U_{\alpha j}U^*_{\beta j}) \sin\, (\Delta m^2_{ij}\frac{L}{2E}) ~~ .
\label{eq15}
\end{eqnarray}
From these expressions, we see that if the mixing matrix $U$  is complex, P($\overline{\nu_\alpha} \ra \overline{\nu_\beta}$) and P($\nu_\alpha \ra \nu_\beta$) will in general differ. Since $\overline{\nu_\alpha} \ra \overline{\nu_\beta}$ is the CP-mirror image of $\nu_\alpha \ra \nu_\beta$, 
P$(\overline{\nu_\alpha} \ra \overline{\nu_\beta}) \neq \mathrm{P}(\nu_\alpha \ra \nu_\beta)$ would be a violation of CP invariance. So far, CP violation has been seen only in the quark sector, so its observation in neutrino oscillation would be most interesting.

With the expressions for P$(\optbar{\nu_\alpha} \ra \optbar{\nu_\beta})$ in hand, let us note several features of neutrino oscillation:
\begin{enumerate}
\item If neutrinos are massless, so that all $\Delta m^2_{ij} = 0$, then, as we see from \Eq{15}, P$(\optbar{\nu_\alpha} \ra \optbar{\nu_\beta})  =  \delta_{\alpha\beta}$. Thus, the observation that neutrinos can change from one flavor to a different one implies neutrino mass. Indeed, it was this observation that led to the conclusion that neutrinos have nonzero masses.

To be sure, every flavor change seen so far has involved neutrinos that travel through matter. Equation~(\ref{eq15}) is for flavor change in vacuum, and does not take into account any interaction between the neutrinos and matter between their source and their detector. Thus, one might ask how we know that the observed flavor changes are not due to flavor-changing interactions between neutrinos and matter, rather than to neutrino masses. In response to this question, two points can be made. First, while the Standard Model of elementary particle physics does not include neutrino masses, it does include a rather well confirmed description of neutrino interactions, and this description states that neutrino interactions with matter do not change flavor. 
Secondly, for at least some of the observed flavor changes, matter effects are expected to be negligible, and there is evidence that for these cases, the flavor-change probability does depend on $L$ and $E$ in the combination $L/E$, as predicted by \Eq{15}. Apart from a constant, $L/E$ is just the proper time that elapses in the rest frame of a neutrino as it travels a distance $L$ with energy $E$. Thus, these flavor changes appear to be an evolution of the neutrino itself over time, rather than a result of interaction with matter.

\item Suppose there is no leptonic mixing. This means that in the decay $W^+ \ra  \overline{\ell_\alpha} + \nu_i$, which as we recall has an amplitude $U_{\alpha i}^*$, the particular charged antilepton $\overline{\ell_\alpha}$ of flavor $\alpha$ is always accompanied by the {\em same} neutrino mass eigenstate $\nu_i$. That is, if $U_{\alpha i}^* \neq 0$, then $U_{\alpha j}$ vanishes for all $j \neq i$. Thus, as we see from \Eq{15}, P$(\optbar{\nu_\alpha} \ra \optbar{\nu_\beta})  =  \delta_{\alpha\beta}$. Hence, the observation that neutrinos can change from one flavor to a different one implies mixing.

\item One can detect neutrino flavor change in two ways. One way is to observe, in a beam of neutrinos which are initially all of flavor $\alpha$, the appearance of neutrinos of a new flavor $\beta$ that is different from the original flavor $\alpha$. This is called an appearance experiment. The other way is to start with a $\nu_\alpha$ beam of known flux, and to observe that some of this known $\nu_\alpha$ flux disappears. This is called a disappearance experiment.

\item Including the so-far-omitted factors of $\hbar$ and $c$, the argument of the oscillatory quantity $\sin^2 [\Delta m^2_{ij} L/4E]$ that appears in \Eq{15} for P$(\optbar{\nu_\alpha} \ra \optbar{\nu_\beta})$ is given by
\beq
\Delta m^2_{ij}\frac{L}{4E} = 1.27 \, \Delta m^2_{ij}(\mathrm{eV}^2) \frac{L\,(\mathrm{km})}{E\, (\mathrm{GeV})} ~~ .
\label{eq16}
\eeq
Now, $\sin^2 [1.27 \, \Delta m^2_{ij}(\mathrm{eV}^2) L\,(\mathrm{km})/E\, (\mathrm{GeV})]$ is appreciable so long as its argument is of order unity or larger. Thus, an experiment with a given $L$ (km) /$E$ (GeV) is sensitive to neutrino squared-mass splittings $\Delta m^2_{ij}(\mathrm{eV}^2)$ all the way down to $\sim [L\,(\mathrm{km})/E\, (\mathrm{GeV}]^{-1}$. For example, an experiment with $L \sim 10^4$ km, the diameter of the earth, and $E \sim 1$ GeV is sensitive to $\Delta m^2_{ij}$ down to $\sim \!10^{-4}$ eV$^2$. As this illustrates, neutrino oscillation provides experimental access to very tiny neutrino masses. It does this by revealing quantum interferences between amplitudes whose relative phases are proportional to neutrino squared-mass differences, but can nevertheless be visibly large if $L/E$ is large enough.

\item As \Eq{15} shows, the probability of flavor change in vacuum oscillates with $L/E$. It is this behavior that has led neutrino flavor change to be called ``neutrino oscillation''.

\item As \Eq{15} indicates, neutrino oscillation probabilities depend only on neutrino squared-mass {\em splittings}, and not on the individual squared neutrino masses themselves. Thus, as illustrated in Fig.~\ref{f2}, oscillation experiments can determine the neutrino squared-mass spectral pattern, but not how far above zero the entire pattern lies.
\begin{figure}[!hbtp]
\begin{center}
\includegraphics[scale=0.7]{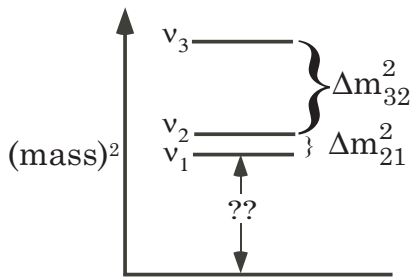}
\caption{A possible neutrino squared-mass spectrum. Oscillation experiments cannot tell us how far above zero the entire spectrum lies. They cannot determine the value of ``??''.} 
\label{f2} 
\end{center}
\end{figure}

\item Neutrino flavor change does not alter the total flux in a neutrino beam, but merely redistributes it among the flavors. Indeed, from \Eq{15} and the unitarity of the $U$ matrix, it follows trivially that
\beq
\sum_\beta \mathrm{P}(\optbar{\nu_\alpha} \ra \optbar{\nu_\beta}) = 1 ~~ ,
\label{eq17}
\eeq
where the sum is to encompass all flavors $\beta$, including the original flavor $\alpha$. \Eq{17} states that the probability that a neutrino changes its flavor, plus the probability that it does not do so, is unity. Hence, flavor change involves no change in total flux. However, some of the flavors $\beta \neq \alpha$ into which a neutrino can transform itself might be {\em sterile} flavors; that is, flavors that do not enjoy normal weak interactions and consequently will not be detected in any feasible detector. If some of the original neutrino flux becomes sterile, then an experiment which measures the total {\em active} neutrino flux---that is, the sum of the $\nu_e,\; \nu_\mu$, and $\nu_\tau$ fluxes---will find it to be less than the original flux.

\item In the literature, treatments of neutrino oscillation frequently assume that the different mass eigenstates $\nu_i$ that contribute coherently to a beam have a common {\em momentum}, rather than the common {\em energy} that we have argued they must have. While the assumption of a common momentum is technically incorrect, it is a harmless error, since, as can easily be shown \cite{r7}, it leads to the same oscillation probabilities as we have found.
\end{enumerate}

An important special case of the not-so-simple formula for P$(\optbar{\nu_\alpha} \ra \optbar{\nu_\beta})$ in \Eq{15} is the case where only two different neutrinos are important. The two-neutrino approximation is a fairly accurate description of a number of experiments. Suppose, then, that only two mass eigenstates, which we shall call $\nu_1$ and $\nu_2$, and two corresponding flavor states, which we shall call $\nu_e$ and $\nu_\mu$, are significant. There is then only one squared-mass splitting, $m^2_2 - m^2_1 \equiv \Delta m^2$. Furthermore, omitting phase factors that can be shown to have no effect on oscillation, the mixing matrix $U$ takes the simple form
\beq\begin{array}{ccc}
     &    \hspace{.55cm} \nu_1 \hspace{.7cm}  \nu_2     \smallskip  \\
U =  & \begin{array}{c}   \nu_e  \\ \nu_\mu  \end{array}  \left[   
	\begin{array}{cc}    \phantom{-}\cos\theta & \sin\theta  \\                        -\sin\theta & \cos\theta   \end{array} \right] ~~ .
	\end{array}
\label{eq18}
\eeq
Here, the symbols above and to the left of the matrix label its columns and rows. The $U$ of \Eq{18} is just a 2$\times$2 rotation matrix, and the rotation angle $\theta$ within it is referred to as the mixing angle. Inserting the $U$ of \Eq{18} and the single $\Delta m^2$ into the general expression for P$(\optbar{\nu_\alpha} \ra \optbar{\nu_\beta})$, \Eq{15}, we immediately find that, for $\beta \neq \alpha$, when only two neutrinos matter,
\beq
\mathrm{P}(\optbar{\nu_\alpha} \ra \optbar{\nu_\beta}) = \sin^2 2\theta \sin^2 (\Delta m^2 \frac{L}{4E}) ~~ .
\label{eq19}
\eeq
In addition, the probability that the neutrino {\em does not} change flavor is, as usual, unity minus the probability that it {\em does} change flavor.

\subsection{Neutrino Flavor Change in Matter}\label{s1.3}

When an accelerator on the earth's surface sends a beam of neutrinos several hundred kilometers to a waiting detector, the beam does not travel through a vacuum, but through earth matter. Coherent forward scattering of the neutrinos in the beam from particles they encounter along the way can then have a large effect. Assuming that the neutrino interactions with matter are the flavor-conserving ones described by the Standard Model, a neutrino in matter can undergo coherent forward scattering from ambient particles in two ways. First, if it is a $\nu_e$---and only if it is a $\nu_e$---it can exchange a $W$ boson with an electron. Coherent forward scattering by electrons via $W$ exchange gives rise to an extra interaction potential energy $V_W$ possessed by electron neutrinos in matter. Clearly, this extra energy from a lowest-order weak interaction will be proportional to $G_F$, the Fermi coupling constant. Equally clearly, this extra energy from $\nu_e-e$ scattering will be proportional to $N_e$, the number of electrons per unit volume. From the Standard Model, we find that 
\beq
V_W = + \sqrt{2}\, G_F\, N_e ~~ ,
\label{eq20}
\eeq
and that this interaction potential energy changes sign if we replace the $\nu_e$ in the beam by $\overline{\nu_e}$.

Secondly, a neutrino in matter can exchange a $Z$ boson with an ambient electron, proton, or neutron. The Standard Model tells us that any flavor of neutrino can do this, and that the amplitude for this $Z$ exchange is flavor independent. The Standard Model also tells us that, at zero momentum transfer, the $Z$ couplings to electron and proton are equal and opposite. Thus, assuming the matter through which our neutrino is traveling is electrically neutral (equal electron and proton densities), the electron and proton contributions to coherent forward neutrino scattering via $Z$ exchange will cancel out. Thus, the $Z$ exchange will give rise to a neutrino-flavor-independent extra interaction potential energy $V_Z$ that depends only on $N_n$, the number of neutrons per unit volume. From the Standard Model, we find that
\beq
V_Z = -\frac{\sqrt{2}}{2}\, G_F\, N_n ~~ ,
\label{eq21}
\eeq
and that, as for $V_W$, this interaction potential energy changes sign if we replace the neutrinos in the beam by antineutrinos.

As already noted, Standard Model interactions do not change neutrino flavor. Thus, unless hypothetical non-Standard-Model flavor-changing interactions are at work, the observation of neutrino flavor change implies neutrino mass and mixing even when the neutrinos are passing through matter.

Neutrino propagation in matter is conveniently treated via a Schr\"{o}dinger laboratory-frame time-evolution equation of the form
\beq
i \frac{\partial}{\partial t} |\nu(t)>\; = \mathcal{H} |\nu(t)> ~~ .
\label{eq22}
\eeq
Here, $|\nu(t)\!>$ is a multi-component neutrino state vector, with one component for each of the possible neutrino flavors. Correspondingly, the Hamiltonian $\mathcal{H}$ is a matrix in flavor space. To illustrate, let us describe the simple case where only two neutrino flavors, say $\nu_e$ and $\nu_\mu$, are important. Then
\beq  
|\nu(t)> \;= \left[   \begin{array}{c}
					f_e (t) \\ f_\mu (t)  \end{array} \right] ~~ ,
\label{eq23}
\eeq
where $f_e (t)$ is the amplitude for the neutrino to be a $\nu_e$ at time $t$, and similarly for $f_\mu (t)$. Correspondingly, $\mathcal{H}$ is a 2$\times$2 matrix in $\nu_e-\nu_\mu$ space.

It is instructive to work out the two-neutrino case first in vacuum, and then in matter. Using \Eq{1} for $|\nu_\alpha >$ in terms of the mass eigenstates, we find that the $\nu_\alpha - \nu_\beta$ matrix element of the vacuum Hamiltonian, $\mathcal{H}_{\mathrm{Vac}}$, is given by
\begin{eqnarray}
 <\nu_\alpha | \mathcal{H}_{\mathrm{Vac}} | \nu_\beta > & = &  <\sum_i U^*_{\alpha i} \nu_i | \mathcal{H}_{\mathrm{Vac}} |\sum_j U^*_{\beta j}\nu_j > \nonumber \\
	& = & \sum_j U_{\alpha j} U^*_{\beta j} \sqrt{p^2 + m_j^2} ~~ .
\label{eq24}
\end{eqnarray}
Here, we are assuming that our neutrino is in a beam of definite momentum $p$, common to all its mass eigenstate components. (As has already been remarked, this assumption, while technically incorrect, will lead us to the correct oscillation probability.) In the last step of \Eq{24}, we have used the fact that $\mathcal{H}_{\mathrm{Vac}} | \nu_j> = E_j|\nu_j>$, where $E_j = \sqrt{p^2 + m_j^2}$ is the energy of mass eigenstate $\nu_j$ at momentum $p$, and the fact that the mass eigenstates of the Hermitean Hamiltonian $\mathcal{H}_{\mathrm{Vac}}$ must be orthogonal.

As we have seen, neutrino flavor change is a quantum interference phenomenon. In this interference, only the {\em relative} phases of the interfering contributions matter. Thus, only the {\em relative} energies of these contributions, which will detemine their relative phases, matter. Consequently, if it is convenient, we may freely subtract from the Hamiltonian $\mathcal{H}$ any multiple of the identity matrix $I$. This will not affect the differences between the eigenvalues of $\mathcal{H}$, and thus it will not affect the predictions of $\mathcal{H}$ for flavor change.

Of course, in the two-neutrino case there are only two mass eigenstates, $\nu_1$ and $\nu_2$, with one splitting $\Delta m^2 \equiv m^2_2 - m^2_1$ between them, and the $U$ matrix is given by \Eq{18}. Inserting this matrix into \Eq{24}, making the highly-relativistic approximation $(p^2 + m^2_j)^{1/2} \cong p + m^2_j/2p$, and subtracting from $\mathcal{H}_{\mathrm{Vac}}$ an irrelevant multiple of the identity matrix, we obtain
\beq
\mathcal{H}_{\mathrm{Vac}} = \frac{\Delta m^2}{4E} \left[
	\begin{array}{cc}
	-\cos 2\theta & \sin 2\theta  \\
	\phantom{-}\sin 2\theta & \cos 2\theta
	\end{array}		\right] ~~ .
\label{eq25}
\eeq
In writing this expression, we have used $p \cong E$, where $E$ is the average energy of the neutrino mass eigenstates in our highly relativistic beam of momentum $p$.

It is easy to confirm that the Hamiltonian $\mathcal{H}_{\mathrm{Vac}}$ of \Eq{25} leads to the same two-neutrino oscillation probability, \Eq{19}, as we have already found by other means. For example, consider the oscillation $\nu_e \ra \nu_\mu$. From the first row of \Eq{18} for $U$,
\beq
|\nu_e> \; = \phantom{-} |\nu_1> \cos\theta + |\nu_2 > \sin \theta ~~ ,
\label{eq26}
\eeq
while from the second row,
\beq
|\nu_\mu> \; = -|\nu_1> \sin\theta + |\nu_2 > \cos \theta ~~ .
\label{eq27}
\eeq
Now, the eigenvalues of $\mathcal{H}_{\mathrm{Vac}}$, \Eq{25}, are
\beq
\lambda_1 = -\frac{\Delta m^2}{4E} ~~ , \;\lambda_2 = +\frac{\Delta m^2}{4E} ~~.
\label{eq28}
\eeq
The corresponding eigenvectors, $|\nu_1>$ and $|\nu_2 >$, are related to $|\nu_e>$ and $|\nu_\mu>$ by Eqs.~(\ref{eq26}) and (\ref{eq27}). Thus, with $\mathcal{H}$ the $\mathcal{H}_{\mathrm{Vac}}$ of \Eq{25}, the Schr\"{o}dinger equation of \Eq{22} implies that if at time $t=0$ we start with a $|\nu_e>$, then after a time $t$ this $|\nu_e>$ will evolve into the state
\beq
|\nu (t)> \;= |\nu_1> e^{+i\frac{\Delta m^2}{4E}t} \cos\theta + |\nu_2 >  e^{-i\frac{\Delta m^2}{4E}t} \sin \theta ~~ .
\label{eq29}
\eeq
The probability P$(\nu_e \ra \nu_\mu)$ that this time-evolved neutrino will be detected as a $\nu_\mu$ is then, from Eqs.~(\ref{eq27}) and (\ref{eq29}),
\begin{eqnarray}
\mathrm{P}(\nu_e \ra \nu_\mu) & = & |<\nu_\mu | \nu(t)>|^2   \nonumber \\
	& = & |\sin\theta\cos\theta (-e^{i\frac{\Delta m^2}{4E}t} + e^{-i\frac{\Delta m^2}{4E}t}) |^2  \nonumber \\
	& = & \sin^2 2\theta \sin^2 (\Delta m^2 \frac{L}{4E}) ~~ .
\label{eq30}
\end{eqnarray}
In the last step, we have replaced our highly-relativistic neutrino's travel time $t$ by its travel distance $L$. The flavor change probability of \Eq{30} does indeed agree with what we found earlier, \Eq{19}.

Let us turn now to neutrino propagation in matter. There, the 2$\times$2 vacuum Hamiltonian $\mathcal{H}_{\mathrm{Vac}}$ is replaced by a matrix $\mathcal{H}_M$ given by
\beq
\mathcal{H}_M = \mathcal{H}_{\mathrm{Vac}} + 
	V_W \left[ \begin{array}{cc} 1 & 0 \\ 0 & 0  \end{array} \right] +
	V_Z \left[ \begin{array}{cc} 1 & 0 \\ 0 & 1  \end{array} \right] ~~.
\label{eq31}
\eeq
Here, the second term on the right-hand side is the contribution from the interaction potential energy caused by $W$ exchange, \Eq{20}. Since this energy affects only $\nu_e$, its contribution is nonvanishing only in the upper left, $\nu_e - \nu_e$, element of $\mathcal{H}_M$. The last term on the right-hand side of \Eq{31} is the contribution from the interaction potential energy caused by $Z$ exchange, \Eq{21}. Since this energy affects all flavors equally, its contribution to $\mathcal{H}_M$ is a multiple of the identity matrix, and consequently can be dropped. Then
\beq
\mathcal{H}_M = \mathcal{H}_{\mathrm{Vac}} + 
	\frac{V_W}{2}\left[ \begin{array}{cc} 1 & 0 \\ 0 & -1 \end{array} \right] +
	\frac{V_W}{2}\left[ \begin{array}{cc} 1 & 0 \\ 0 &  1 \end{array} \right]~~,
\label{eq32}
\eeq
where we have now split the $W$-exchange contribution into a piece that is not proportional to the identity, plus a piece that is proportional to it. Dropping the irrelevant latter piece as well, we have from Eqs.~(\ref{eq25}) and (\ref{eq32})
\beq
\mathcal{H}_M = \frac{\Delta m^2}{4E}  \left[ \begin{array}{cc}
	-(\cos 2\theta - x)  &  \sin 2\theta   \\
	 \sin 2\theta  &  (\cos 2\theta - x)  \end{array}  \right]  ~~ ,
\label{eq33}
\eeq
in which
\beq
x\equiv\frac{V_W /2}{\Delta m^2/4E} = \frac{2\sqrt{2} G_F N_e E}{\Delta m^2} ~~.
\label{eq34}
\eeq
The parameter $x$ is a measure of the importance of the matter effect relative to that of the neutrino squared-mass splitting.

If we define
\beq
\Delta m^2_M \equiv \Delta m^2 \sqrt{\sin^2 2\theta + (\cos 2\theta - x)^2}
\label{eq35}
\eeq
and
\beq
\sin^2 2\theta_M \equiv \frac{\sin^2 2\theta}{\sin^2 2\theta + (\cos 2\theta - x)^2} ~~ ,
\label{eq36}
\eeq
then $\mathcal{H}_M$ can be written as
\beq
\mathcal{H}_M = \frac{\Delta m^2_M}{4E}  \left[ \begin{array}{cc}
	-\cos 2\theta_M  &  \sin 2\theta_M   \\
	 \phantom{-}\sin 2\theta_M  &  \cos 2\theta_M  \end{array}  \right]  ~~ .
\label{eq37}
\eeq
That is, the Hamiltonian in matter, $\mathcal{H}_M$, is identical to its vacuum counterpart, $\mathcal{H}_{\mathrm{Vac}}$, \Eq{25}, except that the vacuum parameters $\Delta m^2$ and $\theta$ are replaced, respectively, by $\Delta m^2_M$ and $\theta_M$.

Needless to say, the eigenstates of $\mathcal{H}_M$ differ from their vacuum counterparts. The splitting between the effective squared-masses of these eigenstates in matter differs from the vacuum splitting $\Delta m^2$, and the effective mixing angle in matter---the angle that determines the $\nu_e,\nu_\mu$ composition of the eigenstates in matter---differs from the vacuum mixing angle $\theta$. Now, all of the physics of neutrino propagation in matter is contained in the matter Hamiltonian $\mathcal{H}_M$. But, according to \Eq{37}, $\mathcal{H}_M$ depends on the parameters $\Delta m^2_M$ and $\theta_M$ in exactly the same way as the vacuum Hamiltonian $\mathcal{H}_{\mathrm{Vac}}$, \Eq{25}, depends on $\Delta m^2$ and $\theta$. Thus, $\Delta m^2_M$ must be the splitting between the effective squared-masses of the eigenstates in matter, and $\theta_M$ must be the effective mixing angle in matter.

In an experiment where an accelerator-generated neutrino beam is sent to a detector that is, say, 1000 km away, the beam passes through earth matter, but does not penetrate very deeply into the earth. The matter density encountered by such a beam en route is very roughly constant. Thus, the electron density $N_e$, hence the parameter $x$, hence the matter Hamiltonian $\mathcal{H}_M$, is roughly position independent, just like the vacuum Hamiltonian $\mathcal{H}_{\mathrm{Vac}}$. Comparing Eqs.~(\ref{eq37}) and (\ref{eq25}), we then see that since $\mathcal{H}_{\mathrm{Vac}}$ leads to the vacuum oscillation probability P$(\nu_e \ra \nu_\mu)$ of \Eq{30}, $\mathcal{H}_M$ must lead to the in-matter oscillation probability
\beq
\mathrm{P}_M(\nu_e \ra \nu_\mu) = \sin^2 2\theta_M \sin^2 (\Delta m^2_M \frac{L}{4E}) ~~ .
\label{eq38}
\eeq
That is, the oscillation probability in matter is the same as in vacuum, except for the replacement of the vacuum parameters $\theta$ and $\Delta m^2$ by their in-matter equivalents.

How large is the earth matter effect, and what are its consequences? To answer this question, we first note from \Eq{34} that the parameter $x$, which measures the relative importance of matter, is proportional to the neutrino energy $E$. To estimate the proportionality constant, let us imagine that we have an accelerator-generated neutrino beam that travels $\sim$1000 km between its source and its detector. The electron density $N_e$ encountered by such a beam will be that of the earth's mantle. The splitting $\Delta m^2$ that will dominate the behavior of such a beam will be the ``atmospheric'' $\Delta m^2$ that also governs the behavior of atmospheric neutrinos, and whose size is approximately $2.4 \times 10^{-3}$eV$^2$ \cite{r8}. Then from \Eq{34}
\beq
|x| \simeq \frac{E}{12\,\mathrm{GeV}} ~~ .
\label{eq39}
\eeq
Thus, in a beam with $E$, say, 2 GeV, the matter effect is modest but not negligible, while in a beam with $E$, say, 20 GeV, the matter effect is very large.

We recall that the splitting $\Delta m^2$ which appears in \Eq{34} is defined as $m^2_2 - m^2_1$, so that, depending on whether $\nu_2$ is heavier or lighter than $\nu_1$, $\Delta m^2$ is positive or negative. We also recall that if the neutrinos, whose propagation in matter we have treated explicitly, are replaced by antineutrinos, then the interaction potential energy $V_W$, which is positive for neutrinos, reverses sign. As a result of these two effects, the sign of $x$, which for neutrinos is given by \Eq{34}, is as summarized in Table~\ref{t1}.
\begin{table}[hbp]
\begin{center}
\caption{The sign of the matter-effect parameter $x$. \label{t1} }
\smallskip
\begin{tabular}{l|c|c}
	Character of	&	$x$ for		&	$x$ for			\\
	the Spectrum	&	Neutrinos	&	Antineutrinos	\\
\hline  
	$\nu_2$ heavier than $\nu_1$	&	+	&	--		\\
	$\nu_2$ lighter than $\nu_1$	&	--	&	+		\\
\hline
\end{tabular}
\end{center}

\end{table}
Now, Eqs.~(\ref{eq35}) and (\ref{eq36}) hold for both neutrinos and antineutrinos, so long as the appropriate value and sign of $x$ are used. These equations show that the effective squared-mass splitting in matter, $\Delta m^2_M$, and the effective mixing angle in matter, $\theta_M$, both depend on the sign of $x$. Thus, since $x$(antineutrinos) = -$x$(neutrinos), $\Delta m^2_M$ and $\theta_M$ will have different values for antineutrinos than they do for neutrinos. That is, there will be an asymmetry between antineutrino oscillation and neutrino oscillation that is induced by matter effects. This asymmetry has nothing to do with genuine CP violation, and will have to be disentangled from the antineutrino-neutrino asymmetry that does come from genuine CP violation in order for us to be able to study the latter phenomenon. However, the antineutrino-neutrino asymmetry coming from matter effects is by no means all bad. If the sign of $\cos 2\theta$ is known, then we can use Eqs.~(\ref{eq35}) and (\ref{eq36}), applied to both neutrinos and antineutrinos, to learn whether $x$(neutrinos) is positive and $x$(antineutrinos) is negative, or the other way around. This, in turn, will tell us whether the neutrino we have called $\nu_2$ is heavier or lighter than the one we have called $\nu_1$. As we shall see in Sec.~\ref{s2.3}, it is hoped that a technique of precisely this kind can tell us whether the three-neutrino spectrum that may actually describe nature has the character shown in Fig.~\ref{f2}, or an inverted character, with the closely-spaced pair at the top, rather than at the bottom.

In principle, matter effects can have very dramatic consequences. From \Eq{36} for the effective mixing angle in matter, $\theta_M$, we see that if the vacuum mixing angle $\theta$ is tiny, with, say, $\sin^2 2\theta = 10^{-3}$, but $x \cong \cos 2\theta$, then $\sin^2 2\theta_M$ can be near or at its maximum possible value, unity. This dramatic amplification of a tiny mixing angle in vacuum into a very large one in matter is the ``resonant'' version of the Mikheyev-Smirnov-Wolfenstein effect \cite{r9}. It used to be thought that this dramatic amplification is actually occurring inside the sun. However, we now know that the solar neutrino mixing angle is already quite large ($\sim 34^\circ$) in vacuum \cite{r10}. Thus, while the effect of solar matter on solar neutrinos is still very significant, it is not quite as dramatic as once thought.
\newcommand{\Dm}[1]{\Delta m^2_{\mathrm {#1}} }

\section{WHAT WE HAVE LEARNED AND THE OPEN QUESTIONS}
 \label{s2}
 
 Now that we have discussed the physics of neutrino oscillation, let us consider what the experimental evidence for oscillation has taught us about neutrinos, and the questions that this evidence has raised.

\subsection{What Have We Learned?}\label{s2.1}

\begin{figure}[!hbtp]
\begin{center}
\includegraphics[scale=0.6]{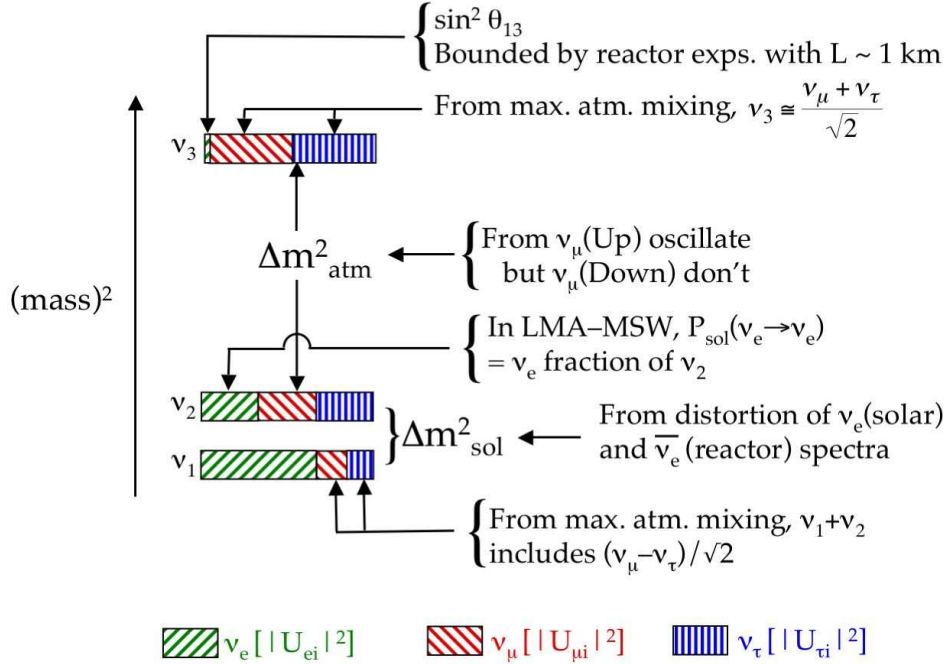}
\caption{A three-neutrino (mass)$^2$ spectrum that accounts for all the neutrino oscillation data except those from LSND. The $\nu_e$ fraction of each mass eigenstate is shown by green right-leaning hatching, the $\nu_\mu$ fraction is shown by red left-leaning hatching, and the $\nu_\tau$ fraction by blue vertical hatching.} 
\label{f3} 
\end{center}
\end{figure}

We do not yet know how many neutrino mass eigenstates there are. Of course, there are at least three, but if the Liquid Scintillator Neutrino Detector (LSND) experiment is confirmed, then there must be more than three, or else our usual assumptions about the neutrinos are wrong in some even more extreme way. On the other hand, if LSND is not confirmed, then Nature may contain only three neutrinos. In that case, the neutrino oscillation data (excluding LSND) tell us that the neutrino squared-mass spectrum is as shown in Fig.~\ref{f3}. 
There are a pair of mass eigenstates, $\nu_1$ and $\nu_2$, separated by the splitting $\Dm{21} \equiv m^2_2 - m^2_1 \equiv \Dm{sol} \cong 8.0 \times 10^{-5}$eV$^2$. This splitting drives the behavior of solar neutrinos. In addition, there is an isolated mass eigenstate, $\nu_3$, separated from $\nu_2$ and $\nu_1$ (``the solar pair'') by the splitting $\Dm{32} \equiv m^2_3 - m^3_2 \equiv \Dm{atm} \cong 2.4 \times 10^{-3}$eV$^2$. This splitting drives the behavior of atmospheric neutrinos.

As explained in Sec.~\ref{s1.1}, each mass eigenstate is a superposition of flavors, and Fig.~\ref{f3} shows its approximate flavor content [see the figure caption].
As mentioned in Sec.~\ref{s1.1}, when mass eigenstate $\nu_i$ interacts and produces a charged lepton, the probability that this charged lepton is, in particular, an electron, is the $\nu_e$ fraction of $\nu_i$, $|U_{ei}|^2$.  The probability that the charged lepton is a muon is the $\nu_\mu$ fraction of $\nu_i$, $|U_{\mu i}|^2$, and the probability that it is a tau is the $\nu_\tau$ fraction, $|U_{\tau i}|^2$.

Fig.~\ref{f3} summarizes how we learned the flavor content of the various mass eigenstates, and the squared-mass splittings between them. With reference to this figure, let us explain how these features of the neutrino spectrum were found, starting with $\nu_3$.

The $\nu_e$ fraction of $\nu_3$ is not known, but is bounded by reactor experiments that had a detector at a distance $L \sim 1$ km from the reactor. Since the (anti)neutrinos emitted by a reactor have an energy $E \sim 3$ GeV, this detector distance made these experiments sensitive to oscillation involving the larger (mass)$^2$ gap, $\Dm{atm} \simeq 2.4 \times 10^{-3}$ eV$^2$, but not to oscillation involving the smaller gap, $\Dm{sol} \simeq 8.0 \times 10^{-5}$ eV$^2$ [cf. \Eq{16} and surrounding text]. 
As a result, these experiments probed the properties of $\nu_3$, the isolated neutrino at one end of the $\Dm{atm}$ gap \cite{r11}. In particular, they probed the $\nu_e$ fraction of $\nu_3$, since the particles emitted by a reactor are $\overline{\nu_e}$. The experiments saw no oscillation of these $\overline{\nu_e}$, whose disappearance they sought, and thereby set a 3$\sigma$ upper bound of $|U_{e3}|^2 < 0.045$ on the $\nu_e$ fraction of $\nu_3$ \cite{r12}.

One hears a lot of discussion of a leptonic mixing angle called $\theta_{13}$. This angle is so defined that $|U_{e3}|^2 = \sin^2 \theta_{13}$. Thus, $\theta_{13}$ is a measure of the smallness of the $\nu_e$ part of $\nu_3$.

Apart from this small $\nu_e$ piece, $\nu_3$ is of $\nu_\mu$ and $\nu_\tau$ flavor. Now, the oscillation of atmospheric muon neutrinos is observed to be dominated by $\nu_\mu \ra \nu_\tau$, with a $\nu_\mu - \nu_\tau$ mixing angle that is very large. The best fit for this angle is maximal mixing: 45$^\circ$. 
This atmospheric mixing angle will be reflected in the flavor content of $\nu_3$, since $\nu_3$ is at one end of the splitting $\Dm{atm}$ that drives atmospheric neutrino oscillation. If the angle is truly maximal, then, apart from its small $\nu_e$ component, $\nu_3$ is simply $(\nu_\mu + \nu_\tau) / \sqrt{2}$. This mimics the behavior of the neutral $K$ meson system. There, apart from a small CP violation, the mixing of $K^0$ and $\overline{K^0}$ is maximal, with the consequence that $K_S = (K^0 + \overline{K^0}) /\sqrt{2}$.

It is found that 1-GeV upward-going atmospheric neutrinos, which originate in the atmosphere on the far side of the earth from their detector, and hence  travel $\sim 10^4$ km---the diameter of the earth---to reach the detector,  oscillate. In contrast, 1-GeV downward-going neutrinos, which originate in the atmosphere directly above their detector, and thus travel only $\sim 10$ km---the thickness of the atmosphere---to reach the detector, do not oscillate. From these facts, and \Eq{16}, it follows that $10^{-4}$eV$^2 \lsim \Dm{atm} \lsim 10^{-2}$eV$^2$. Of course, the detailed experiments pin down $\Dm{atm}$ much better than that.

It is well established that the change of flavor of solar neutrinos is caused by the Large Mixing Angle version of the Mikheyev-Smirnov-Wolfenstein effect (LMA-MSW). As will be explained shortly, according to LMA-MSW, the probability P$_{\mathrm{sol}}(\nu_e \ra \nu_e )$ that solar neutrinos, all of which are born $\nu_e$, will still be $\nu_e$ when they arrive at the earth, is to a good approximation just the $\nu_e$ fraction of the mass eigenstate $\nu_2$. 
Experimentally, P$_{\mathrm{sol}}(\nu_e \ra \nu_e )$ is found to be approximately 1/3 \cite{r10}. Therefore, the $\nu_e$ fraction of $\nu_2$, $|U_{e2}|^2$, is $\sim$ 1/3. From the unitarity of the $U$ matrix, $|U_{e1}|^2 + |U_{e2}|^2 + |U_{e3}|^2 = 1$. Thus, since $|U_{e3}|^2$ is small and $|U_{e2}|^2 \simeq$ 1/3, $|U_{e1}|^2$, the $\nu_e$ fraction of $\nu_1$, is approximately 2/3.

If the atmospheric $\nu_\mu - \nu_\tau$ mixing is maximal, then, together, $\nu_1$ and  $\nu_2$ contain the neutrino $(\nu_\mu - \nu_\tau) / \sqrt{2}$, the 50-50 $\nu_\mu - \nu_\tau$ mixture that is orthogonal to the one occurring in $\nu_3$. Then, $\nu_1,\; \nu_2$, and $\nu_3$ all contain $\nu_\mu$ and $\nu_\tau$ in equal proportion.

The solar (mass)$^2$ splitting, $\Dm{sol}$, has been determined from the observed energy dependence of solar neutrino flavor change, and especially from the observed energy dependence of reactor $\overline{\nu_e}$ disappearance as studied by the KamLAND reactor experiment \cite{r14}. 
Unlike the reactor experiments alluded to earlier, KamLAND has a detector situated, not 1 km, but on average 180 km away from reactors that are its $\overline{\nu_e}$ sources. With this greatly increased source-to-detector distance, KamLAND is sensitive to oscillation involving the small splitting $\Dm{sol}$. 

While the flavor content pictured in Fig.~\ref{f3} tells us the magnitudes $|U_{\alpha i}|^2$, it does not show the signs or phases of the $U$ matrix elements. To discuss these, we need to look directly at $U$ itself. It is illuminating to write $U$ in the form
\begin{eqnarray}
&& \hspace{.4cm} \mathrm{Atmospheric} \hspace{1.3cm} \mbox{Cross-Mixing} \hspace{2.0cm} \mathrm{Solar} \hspace{1.7cm} \parbox{2.45cm}{Majorana CP-violating phases}    \nonumber \\
U&=&	\left[ \begin{array}{ccc}
	1 & 0 & 0 \\ 0 & \phantom{-}c_{23} & s_{23} \\ 0 & -s_{23} & c_{23}
			\end{array}	\right] \times
			\left[ \begin{array}{ccc}
	\phantom{-}c_{13} & 0 & s_{13}e^{-i\delta} \\ 0 & 1 & 0 \\  -s_{13}e^{i\delta} & 0 & c_{13}
			\end{array}	\right] \times
			\left[ \begin{array}{ccc}
	\phantom{-}c_{12} & s_{12} & 0 \\ -s_{12} &c_{12} & 0 \\ 0 & 0 & 1
			\end{array}	\right] \times
			\left[ \begin{array}{ccc}
	e^{i\alpha_1 /2} & 0 & 0 \\ 0 &e^{i\alpha_2 /2} & 0 \\ 0 & 0 & 1
			\end{array}	\right]  ~~ .
\label{eq40}
\end{eqnarray}
Here, $c_{ij} \equiv \cos \theta_{ij}$ and $s_{ij} \equiv \sin \theta_{ij}$, where $\theta_{ij}$ is a mixing angle, and $\delta, \; \alpha_1$, and $\alpha_2$ are CP-violating phases.

In the decomposition of $U$ in \Eq{40}, the first, Atmospheric, factor matrix dominates the mixing exhibited by the atmospheric neutrinos. This factor depends on a mixing angle $\theta_{23}$ that is approximately the atmospheric mixing angle $\theta_{\mathrm{atm}}$ determined when the atmospheric $\nu_\mu \ra \nu_\tau$ oscillation is described by an approximate two-neutrino oscillation formula of the form of \Eq{19}. At 90\% confidence level, $37^\circ \leq \theta_{\mathrm{atm}} \leq 53^\circ$ \cite{r8}. As already mentioned, this mixing is very large. The value of $\theta_{\mathrm{atm}}$ that fits the data best is 45$^\circ$: maximal mixing \cite{r8}.

The third, Solar, matrix on the right-hand side of \Eq{40} dominates the mixing of solar neutrinos. The mixing angle in this matrix, $\theta_{12}$, is approximately the solar mixing angle $\theta_{\mathrm{sol}}$ determined by approximating the solar neutrino problem as a two-neutrino problem. This is a very good approximation, since the solar neutrinos are born as $\nu_e$, and $|U_{e3}|^2$, the $\nu_3$ fraction of $\nu_e$, is quite small. 
Thus, $\nu_e$ is approximately a mixture of just $\nu_1$ and $\nu_2$, and it will remain so as it propagates. Experimentally, $\theta_{\mathrm{sol}} = (33.9 {+2.4 \atop -2.2})$ degrees \cite{r10} . Although this mixing angle is not maximal, as $\theta_{\mathrm{atm}}$ might be, it is still very large. It is interesting that two of the leptonic mixing angles are large, while all of the quark mixing angles are small. Is there a clue to the physics underlying fermion mass and mixing in this disparity?

The second, Cross-Mixing, matrix in \Eq{40} involves the small mixing angle $\theta_{13}$, which measures the small $\nu_e$ part of $\nu_3$. This matrix also may involve a CP-violating phase $\delta$. A nonvanishing $\delta$ will in general lead to a CP-volating difference between the probabilities for the CP-mirror-image oscillations $\nu_\alpha \ra \nu_\beta$ and $\overline{\nu_\alpha} \ra \overline{\nu_\beta}$. We note from \Eq{40}, however, that $\delta$ enters the mixing matrix $U$ only in combination with $\sin \theta_{13}$. Thus, the size of the CP-violating difference $P(\nu_\alpha \ra \nu_\beta) - P(\overline{\nu_\alpha} \ra \overline{\nu_\beta})$ will depend on $\theta_{13}$. Consequently, the power of the facilities we will need in order to see this CP-violating difference will also depend, at least roughly, on $\theta_{13}$. At present, we know only that, at 3$\sigma$, $\theta_{13} < 12^\circ$ \cite{r12}. Clearly, demonstrating that $\theta_{13}$ is nonvanishing, and determining its approximate size, are rather important.

The final matrix in \Eq{40} contains so-called ``Majorana'' CP-violating phases that have no analogue in the quark sector, and that have physical effects only if the neutrinos are their own antiparticles. In particular, these phases have no effect on neutrino oscillation, which is completely insensitive to whether neutrinos are their own antiparticles.

The reader may wonder why the CP-violating phase $\delta$ enters $U$ in combination with $\theta_{13}$, rather than $\theta_{12}$ or $\theta_{23}$.  Doesn't the $U$ matrix decomposition of \Eq{40} treat all three mixing angles symmetrically? The answer to this puzzle is that if one multiplies the matrices in \Eq{40} to obtain the detailed expression for $U$, and then uses one's freedom to rephase elements of $U$ by phase-redefining any neutrino or charged lepton, then one can completely remove the phase $\delta$ from $U$ if {\em any} of the three mixing angles vanishes. 
That is, for $\delta$ to have physical consequences, all three of the mixing angles must be non-zero. However, we already know that two of them, $\theta_{12}$ and $\theta_{23}$, are very far from zero. Therefore, we have chosen to write $U$ in a form that emphasizes the fact that the CP-violating consequences of $\delta$ will disappear if the remaining mixing angle $\theta_{13}$---the only mixing angle we don't know---should vanish.

\begin{figure}[!hbtp]
\begin{minipage}[b]{8.0cm}
\includegraphics[scale=0.5]{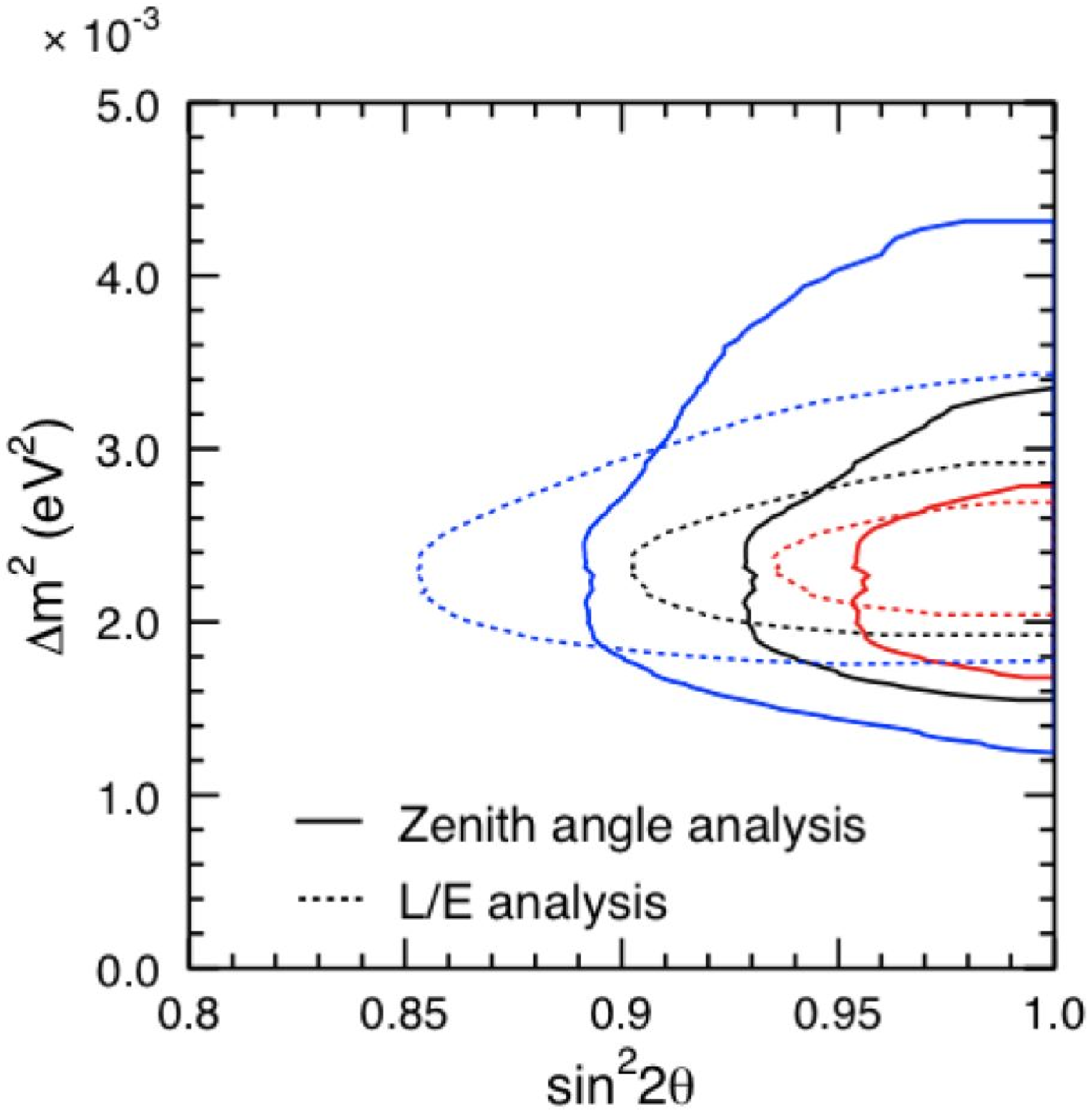}
\caption{Allowed region for the atmospheric $\Dm{}$ and mixing angle from the Super-Kamiokande experiment \cite{r8}. The solid and dashed curves come from two different data analyses. For each analysis, the innermost, middle, and outermost curves bound the 68, 90, and 99\% confidence level regions, respectively.} 
\label{f4}
\end{minipage}    \hfill     \begin{minipage}[b]{8cm}
\includegraphics[scale=0.5]{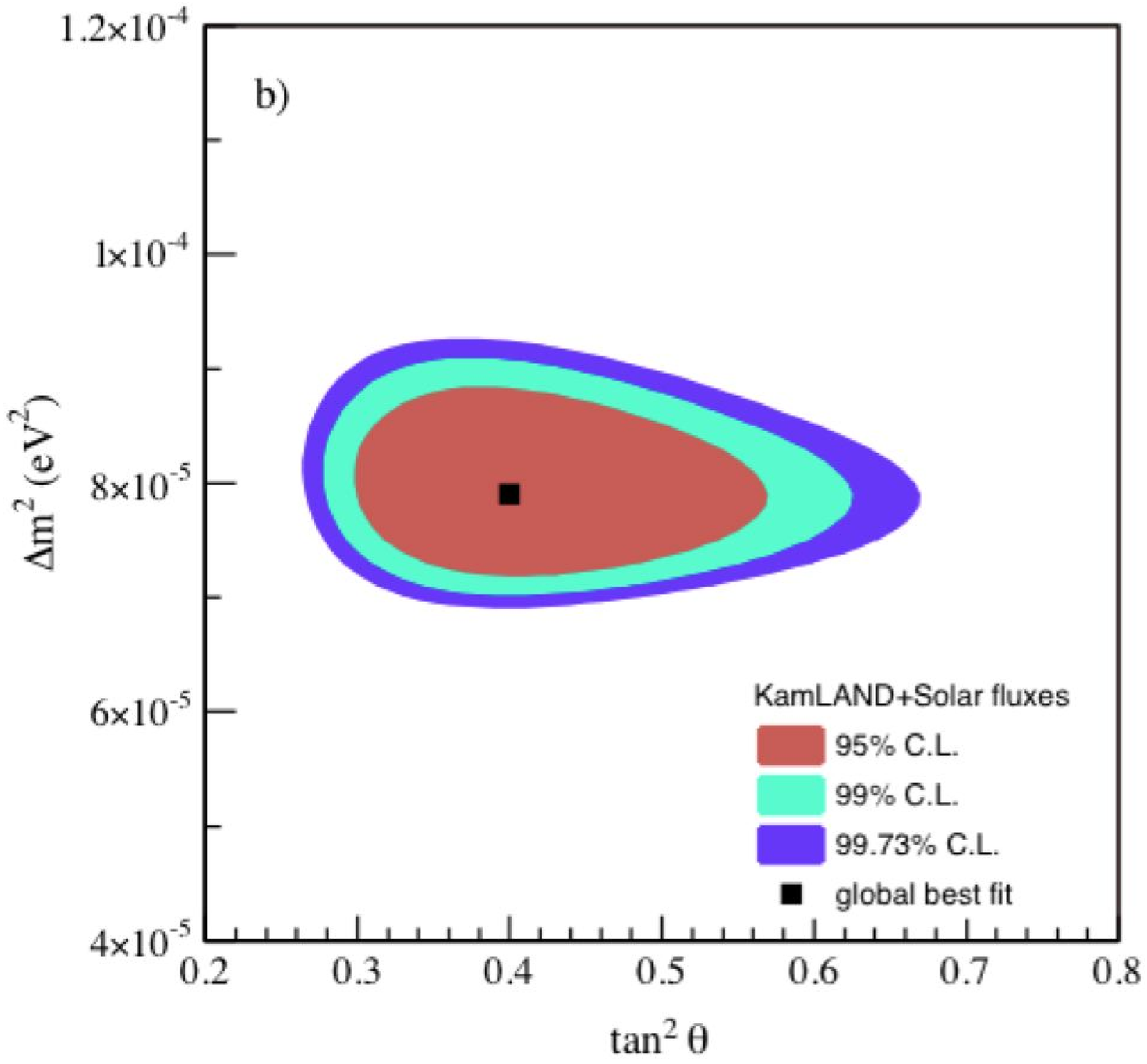}
\caption{Allowed region for the solar $\Dm{}$ and mixing angle, from a KamLAND analysis of KamLAND and solar neutrino data \cite{r14}.} 
\vspace{1.3cm}
\label{f5}
\end{minipage}
\end{figure}

Figures \ref{f4} and \ref{f5} show graphically the allowed regions for the atmospheric and solar mixing angles and (mass)$^2$ splittings.

\subsection{The Large Mixing Angle MSW Effect}\label{s2.2}

In Sec.~\ref{s1.3}, we discussed neutrino flavor change in matter. Let us now consider how this works in the case of the solar neutrinos, which encounter a lot of matter as they travel from the core of the sun, where they are born, to its outer edge. As mentioned previously, it is now well established that the change of flavor of these neutrinos is governed by the Large Mixing Angle MSW Effect. As we saw from \Eq{34}, the importance of matter effects grows with neutrino energy. 
In the sun, this has the consequence that for the neutrinos produced in $^8$B decay, which are at the high end of the solar neutrino energy spectrum, the influence of solar matter is quite significant. However, for the neutrinos produced by $pp$ fusion, which have an energy an order of magnitude lower, the effect of matter is tiny. Since it is the $^8$B neutrinos that have been most extensively studied, let us focus on them.

As we have already noted, solar neutrino propagation is approximately a two-neutrino phenomenon. Solar neutrinos are mixtures of $\nu_1$ and $\nu_2$, with a $\nu_3$ component that may be neglected. The two mass eigenstates $\nu_1$ and $\nu_2$ make up two flavor states: $\nu_e$ and a state $\nu_x$ that is a coherent linear combination of $\nu_\mu$ and $\nu_\tau$. The precise $\nu_\mu - \nu_\tau$ composition of $\nu_x$ depends on the $\nu_\mu - \nu_\tau$ mixing angle $\theta_{\mathrm{atm}}$ measured in atmospheric neutrino oscillation. For $\theta_{\mathrm{atm}} = 45^\circ,\; \nu_x = (\nu_\mu - \nu_\tau) / \sqrt{2}$, a 50-50 mixture. Solar neutrino flavor change is the process $\nu_e \ra \nu_x$.

Since solar neutrino propagation involves only two neutrinos, we may describe it by the two-neutrino in-matter Hamiltonian $\mathcal{H}_M$ of \Eq{31}. Dropping the irrelevant $Z$ exchange term as before, and using Eqs.~(\ref{eq25}) and (\ref{eq20}), we have
\beq
\mathcal{H}_M = \frac{\Dm{sol}}{4E}
	\left[  \begin{array}{rr}
		-\cos 2\theta_{\mathrm{sol}}  & \sin 2\theta_{\mathrm{sol}} \\ 
		\phantom{-}\sin 2\theta_{\mathrm{sol}} & \cos 2\theta_{\mathrm{sol}}
			\end{array}  \right]
	+ \sqrt{2} G_F N_e
	\left[	\begin{array}{cc}
		1 & 0  \\ 0 & 0
			\end{array}  \right] ~~.
\label{eq41}
\eeq
The matrices in this equation are in the $\nu_e - \nu_x$ flavor basis, and we have specialized to the case of solar neutrinos, where the relevant vacuum squared-mass splitting is $\Dm{sol}$ and the relevant vacuum mixing angle is $\theta_{\mathrm{sol}}$.

At the center of the sun, where the solar neutrinos are created, the coefficient of the second, matter-interaction term in the $\mathcal{H}_M$ of \Eq{41} is
\beq
\sqrt{2} G_F N_e \approx 0.75 \times 10^{-5} \mathrm{eV}^2 / \mathrm{MeV} ~~.
\label{eq42}
\eeq
By comparison, since $\Dm{sol} \simeq 8 \times 10^{-5}$eV$^2$, at a typical $^8$B neutrino energy $E$ of $\sim$8 MeV, the coefficient of the first, vacuum term in this $\mathcal{H}_M$ is
\beq
\frac{\Dm{sol}}{4E} \approx 0.25 \times 10^{-5} \mathrm{eV}^2 / \mathrm{MeV} ~~.
\label{eq43}
\eeq
Thus, at $r = 0$ (where $r$ is the distance from the center of the sun), the interaction term in $\mathcal{H}_M$ dominates, at least to some extent. Let us then, as a first approximation, neglect the vacuum term in $\mathcal{H}_M (r=0)$. As we see from \Eq{41}, $\mathcal{H}_M (r=0)$ is then diagonal. This means that a $\nu_e$ born at $r=0$ is an eigenstate of $\mathcal{H}_M (r=0)$. Since, from \Eq{41}, the $\nu_e$ eigenvalue of (the approximate) $\mathcal{H}_M (r=0)$ is $\sqrt{2} G_F\, N_e$, while the other eigenvalue is zero, our $\nu_e$ is born in the higher-energy eigenstate of $\mathcal{H}_M (r=0)$.

Under the conditions where the Large Mixing Angle MSW effect occurs, the propagation of a $^8$B neutrino from $r \simeq 0$ to the outer edge of the sun is adiabatic. That is, the electron density $N_e$ changes slowly enough with $r$ that the Schr\"{o}dinger equation, \Eq{22}, in which $\mathcal{H}_M$ acts may be solved for one radius $r$ at at time, and then the solutions may be patched together. Our neutrino will propagate outward as the slowly-changing eigenstate of the slowly changing $\mathcal{H}_M (r)$. 
Furthermore, it can easily be shown that, as $N_e$ changes with $r$, the eigenvalues of $\mathcal{H}_M (r)$, \Eq{41}, never cross. Thus, our neutrino, having begun as the higher-energy eigenstate of $\mathcal{H}_M (r = 0)$, will emerge from the outer edge of the sun as the higher-energy eigenstate of $\mathcal{H}_M (r =$ radius of sun). But $\mathcal{H}_M (r =$ radius of sun) is just the vacuum Hamiltonian $\mathcal{H}_{\mathrm{Vac}}$, since at the outer edge of the sun, the electron density $N_e$ has fallen to zero. Thus, our neutrino emerges from the sun as the heavier mass eigenstate of the two-neutrino vacuum Hamiltonian. This is the mass eigenstate we have called $\nu_2$ in Fig.~\ref{f3}.

Being a mass eigenstate, our $\nu_2$ is an ordinary particle, just like an electron, and it will propagate from the outer edge of the sun to the surface of the earth without changing into anything else. The probability that when it interacts here on earth in a detector, it will do so like a $\nu_e$, producing an electron, is then just the $\nu_e$ fraction of $\nu_2$, $|U_{e2}|^2$. The Sudbury Neutrino Observatory (SNO) has found this probability to be approximately one-third \cite{r10}. Therefore, $|U_{e2}|^2 \simeq 1/3$. 
(In the approximation where we treat the solar neutrinos neglecting $\nu_3$, the $U$ matrix is given by \Eq{18}, with $\theta$ the solar mixing angle $\theta_{\mathrm{sol}}$, and with $\nu_\mu$ replaced by $\nu_x$. Then $|U_{e2}|^2 = \sin^2  \theta_{\mathrm{sol}}$, so the SNO results imply that $\sin^2  \theta_{\mathrm{sol}}\simeq 1/3$.)

This is how the Large Mixing Angle MSW effect works, at least to a good approximation. It should be noted that, while this effect causes a change in the flavor content of a neutrino, there is no sinusoidal oscillation of the sort that characterizes the flavor change in vacuum described by \Eq{15} or \Eq{19}. 
In LMA-MSW, a $\nu_e$ gradually evolves into a $\nu_2$ within the sun, without anything sinusoidally oscillating. Then, this $\nu_2$ propagates to the earth without changing further, so that again nothing is oscillating. To be sure, a neutrino born as a $\nu_e$ has changed into a mixture of the flavors $\nu_e$ and $\nu_x$. But this flavor change has not involved any undulation.

\subsection{What Would We Like To Find Out?}\label{s2.3}

Having reviewed what we have learned about the neutrinos from existing data, let us now ask what we would like to find out through future experiments. Let us consider just some of the most interesting open questions.

\begin{itemize}
\item How many neutrino species are there? Are there sterile neutrinos?
\end{itemize}

As has already been noted, we do not know how many neutrino mass eigenstates there are, with the LSND result suggesting that there are more than three. If there are indeed more than three mass eigenstates, then there are linear combinations of them that do not couple to the $W$ or $Z$ bosons. 
These linear combinations---so-called sterile neutrinos---do not participate in any of the known forces except gravity. Thus, confirmation of LSND would have far-reaching consquences. The MiniBooNE experiment, presently taking data, aims to confirm or refute LSND.

\begin{itemize}
\item What are the masses of the neutrino mass eigenstates?
\end{itemize}

This question has at least two parts. First, we know, within errors, the values of the atmospheric and solar splittings, $\Dm{atm}$ and $\Dm{sol}$. However, we do not know whether the solar pair, $\nu_1$ and $\nu_2$, separated by the smaller of the two splittings, $\Dm{sol}$, actually lies at the bottom of the spectrum, as in Fig.~\ref{f3}, or at the top. If this pair is at the bottom, the spectrum is called ``normal'', and if it is at the top, the spectrum is called ``inverted''. 
Whether the spectrum is normal or inverted is an interesting question, because generically, grand unified theories (GUTS) favor a normal spectrum \cite {r16}. This may be understood by remembering that GUTS relate the leptons to the quarks, and the quarks have normal spectra, with the most closely spaced quark pair at the bottom of the spectrum, not at the top. An inverted neutrino spectrum would be quite un-quark-like, and would probably involve a lepton symmetry, with no quark analogue, as the source of the near degeneracy of $\nu_1$ and $\nu_2$ at the top of the spectrum.

To determine whether the spectrum is normal or inverted, one may study the effect of earth matter on $\nu_\mu \ra \nu_e$ and $\overline{\nu_\mu} \ra \overline{\nu_e}$ oscillations. Using accelerator-generated $\nu_\mu$ and $\overline{\nu_\mu}$ beams, one will carry out a long-baseline experiment with an $L/E$ that makes it sensitive to the atmospheric mass gap, $\Dm{atm} \equiv \Dm{32}$. The latter quantity is positive for a normal spectrum, but negative for an inverted one. 
By studying oscillations into a $\nu_e$ or $\overline{\nu_e}$, one will gain sensitivity to the $W$-exchange interaction betwen these particles and electrons in the earth matter through which the beam passes. As previously noted, this interaction leads to the extra interaction potential energy $V_W = + \sqrt{2} G_F N_e$ for a $\nu_e$ [cf. \Eq{20}], and to an equal but opposite extra energy for a $\overline{\nu_e}$. 
The signs of these extra energies are known from the Standard Model. By probing oscillations in which the quantity whose sign we wish to determine, $\Dm{32}$, interferes with the $W$-exchange-induced extra $\nu_e$ or $\overline{\nu_e}$ energy, whose sign we already know, we can determine the sign of $\Dm{32}$.

Since the experiment will involve $\Dm{atm}$ and focus on a $\nu_e$ or $\overline{\nu_e}$ final state, it will require that the $\nu_e$ fraction of $\nu_3$, the neutrino at one end of $\Dm{atm}$, be nonvanishing. Of course, this fraction is $\sin^2 \theta_{13}$. Thus, not only CP violation in neutrino oscillation, but also one's ability to tell whether the spectrum is normal or inverted by using matter effects, depends on a nonvanishing $\theta_{13}$ \cite{r17}.

In earth matter, the vacuum mixing angle $\theta_{13}$ is replaced by an effective mixing angle $\theta_M$ given by \Eq{36}, with $\theta$ taken to be $\theta_{13}$. For a neutrino beam, the parameter $x$ in \Eq{36} is defined by \Eq{34}, with $\Dm{}$ now taken to be $\Dm{atm} \equiv \Dm{32}$. For an antineutrino beam, $x$ reverses sign. Note that $x$ embodies the interference between $\Dm{32}$ and $V_W$, enabling us to determine the unknown sign of the first from the known sign of the second, as just described.

From \Eq{39} and the experimental knowledge that $\cos 2\theta_{13} > 0.91$, we see that for neutrino beam energies $E \lsim 2$ GeV, we may expand the denominator of \Eq{36} to obtain
\beq
\sin^2 2 \theta_M \simeq \sin^2 2\theta_{13} [1\, \poptm\, \mathcal{S} \frac{E}{6\, \mathrm{GeV}} ] ~~ .
\label{eq46}
\eeq
Here, $\mathcal{S}$ is the sign of $\Dm{32}$ [see \Eq{34}], and the positive (negative) sign on the right-hand side of the equation is for a neutrino (antineutrino) beam.

At oscillation maximum, the $\nu_\mu \ra \nu_e$ and $\overline{\nu_\mu} \ra \overline{\nu_e}$ oscillation probabilities will be proportional to the $\sin^2 2\theta_M$ given by \Eq{46}. Thus, at oscillation maximum,
\begin{displaymath}    
\frac{P(\nu_\mu \ra \nu_e)}{P(\overline{\nu_\mu} \ra \overline{\nu_e})}   \hspace{.5cm} \mbox{is} \hspace{.5cm}
			\left\{ \begin{array}{c}
	>1 ~~; ~~\mbox{Normal Spectrum}\;\; (\mathcal{S} = +1)  \\
	<1 ~~; ~~\mbox{Inverted Spectrum}\; (\mathcal{S} = -1)  \\
			\end{array} \right.  ~~ .
\end{displaymath}
As \Eq{46} shows, at, say, $E \simeq$ 2 GeV, this ratio can deviate from unity quite substantially.

A second aspect of the question ``What are the masses of the neutrino mass eigenstates?'' is the issue of the absolute scale of neutrino mass. How high above the zero of mass does the entire spectral pattern depicted in Figs.~\ref{f2} and \ref{f3} lie? As already noted, neutrino flavor change, which depends only on  squared-mass {\em splittings}, and not on individual masses, cannot answer this question. 
To be sure, flavor change does provide a lower bound on the mass of the heaviest mass eigenstate. Namely, since the mass of the lightest mass eigenstate is not below zero, the mass of the heaviest one cannot be less than $\sqrt{\Dm{atm}} \simeq$ 0.04 eV.

Approaches to going beyond a lower bound to an actual numerical value include the study of tritium $\beta$ decay, neutrinoless nuclear double $\beta$ decay, and cosmology. Cosmology has already provided an interesting {\em upper} bound. Cosmological data plus several cosmological assumptions imply that 
\beq
\sum_i m_i < (0.4 - 1.0) \, \mathrm{eV} ~~,
\label{eq47}
\eeq
where $m_i$ is the mass of mass eigenstate $\nu_i$, and the sum runs over all the mass eigenstates \cite{r18}. Within the context of a three-neutrino spectrum with the observed squared-mass splittings, this bound implies that the mass of the heaviest $\nu_i$ is less than (0.2 -- 0.4) eV. Thus, combining the flavor change and cosmological information,
\beq
0.04\, \mathrm{eV} < \mbox{Mass\,[Heaviest $\nu_i$]} < (0.2-0.4)\, \mathrm{eV} ~~.
\label{eq48}
\eeq
This is already an interesting indication of the scale of neutrino mass. However, we would like to learn more about the reliability of the cosmological assumptions, and we would like to aim for, not just bounds, but a numerical value.

\begin{itemize}
\item How large is $\theta_{13}$?
\end{itemize}

As we have seen, both CP violation and our ability to tell whether the mass spectrum is normal or inverted depend on $\theta_{13}$. Thus, we would like to know at least the approximate value of this angle. To see how we may learn that, let us recall that $\sin^2\theta_{13}$ is the small $\nu_e$ fraction of the isolated mass eigenstate $\nu_3$ [see Fig.~\ref{f3}]. 
This mass eigenstate is at one end of the atmospheric splitting, $\Dm{atm}$. Thus, to measure $\theta_{13}$, we must do an oscillation experiment whose $L/E$ makes it sensitive to $\Dm{atm}$, and that involves a $\nu_e$, either as the initial or final flavor. Two complementary approaches are being pursued. In the first, one will seek to observe the disappearance of reactor $\overline{\nu_e}$, which are particles of the required electron flavor, while they travel a distance $L \sim$ 1.5 km. 
Since reactor $\overline{\nu_e}$  typically have an energy $E \sim$ 3 MeV, $L/E$  will be $\sim$\,500 km/GeV, making the experiment sensitive, according to \Eq{16}, to the atmospheric splitting $\Dm{atm} \simeq 2.4 \times 10^{-3}$ eV$^2$. In the second approach, one will seek to observe the appearance of $\nu_e$, neutrinos of the required electron flavor, in an accelerator $\nu_\mu$ beam whose energy $E$ is of order 1 GeV, and which travels a distance $L$ of several hundred kilometers from source to detector. The $L/E$ of this experiment will be comparable to that of the reactor experiment.

It can easily be shown that the reactor experiment is sensitive to $\theta_{13}$, and $\theta_{13}$ alone, while the accelerator experiment is sensitive to $\theta_{13}$, $\theta_{23}$, the CP-violating phase $\delta$, and the normal or inverted character of the mass spectrum. Thus, the accelerator approach has the advantage of providing access to several neutrino properties that we wish to learn. However, these properties will have to be disentangled from one another in the data. This disentanglement will be facilitated by a clean measurement of $\theta_{13}$ from a reactor experiment.

\begin{itemize}
\item Are neutrinos their own antiparticles?
\end{itemize}

Alone among the fundamental fermionic constituents of matter, neutrinos might be their own antiparticles. A quark or charged lepton cannot be its own antiparticle, because it is electrically charged, and its antiparticle has the opposite electric charge. However, neutrinos are not electrically charged and, as far as we know, they do not carry any other conserved, charge-like quantity either. It might be thought that there is a conserved charge-like lepton number $L$, defined by
\beq
L(\nu) = L(\ell) = -L(\bar{\nu}) = -L(\bar{\ell}) = 1 ~~ ,
\label{eq49}
\eeq
that distinguishes neutrinos $\nu$ and charged leptons $\ell$ from their antiparticles. However, there is absolutely no evidence that any such conserved quantum number exists. If it does not exist, then there is nothing to distinguish a neutrino from its antiparticle. Then each neutrino mass eigenstate $\nu_i$ is identical to its antiparticle $\overline{\nu_i}$, making the neutrinos very different from the charged leptons and the quarks. 

A neutrino that is identical to its antiparticle is referred to as a Majorana particle, while one that is not is referred to as a Dirac particle.

Many theorists believe that, quite likely, the lepton number $L$ defined by \Eq{49} is not conserved, so that neutrinos are Majorana particles. One reason for this widely-held belief is the character of possible neutrino mass terms in extensions of the Standard Model (SM) \cite{r19}. The SM can be defined by the fields it contains, and by some general principles, such as weak isospin invariance and renormalizability. It is observed that anything allowed by the field content of the pre-neutrino-mass, original version of the SM, and allowed by the general SM principles, occurs in Nature. 

The original version of the SM, which omits neutrino masses and contains no chirally right-handed neutrino fields, $\nu_R$, but only left-handed ones, $\nu_L$, conserves $L$. However, now that we know neutrinos have masses, we must extend the SM to accommodate them. Suppose that we try to do this in a way that will preserve the conservation of $L$. Then, for a neutrino $\nu$, we add to the SM Lagrangian a so-called ``Dirac mass term'' of the form
\beq
\mathcal{L}_D = -m_D\, \overline{\nu_L}\, \nu_R + \mathrm{h.c.} ~~.
\label{eq50}
\eeq
Here, $m_D$ is a constant, and $\nu_R$ is a right-handed neutrino field that we are forced to add to the SM in order to construct the Dirac mass term. This mass term is of the same form as those that give masses to the charged leptons and quarks. It does not mix neutrinos and antineutrinos, and consequently conserves $L$.

In the SM, left-handed fermion fields belong to weak-isospin doublets, but right-handed ones are isospin singlets. Thus, the newly-introduced $\nu_R$ will carry no weak isospin. Hence, once $\nu_R$ is present, all the SM principles, notably including weak-isospin conservation, allow the occurrence of the ``Majorana mass term''
\beq
\mathcal{L}_M = -m_M\, \overline{\nu_R^c}\, \nu_R + \mathrm{h.c.} ~~.
\label{eq51}
\eeq
Here, $m_M$ is a constant, and $\nu_R^c$ is the charge conjugate of $\nu_R$. Since $\nu_R$ carries no weak isospin, neither does $\nu_R^c$, so $\mathcal{L}_M$ is fully weak-isospin conserving. However, any Majorana mass term of this form turns a $\nu$ into a $\bar{\nu}$, and a $\bar{\nu}$ into a $\nu$. Thus, it clearly does not conserve $L$.

The neutrino $\nu$ appearing in the mass terms of Eqs.~(\ref{eq50}) and (\ref{eq51}) is not a mass eigenstate, but one of the underlying states in terms of which the model is written. Once the $\nu \leftrightarrow \bar{\nu}$ mixing, hence $L$-nonconserving, Majorana mass term of \Eq{51} is present, nothing can distinguish a neutrino mass eigenstate from its antiparticle, so that diagonalization of the model will yield mass eigenstates $\nu_i$ that are Majorana particles.

Nature has proved to contain (except perhaps in the Higgs sector) everything allowed by the general principles of the SM and the field content of its original version, which omits neutrino masses. It is then natural to expect that she also contains everything allowed by the principles of the SM and the field content of its extended version, which includes neutrino masses. If so, then Nature contains Majorana mass terms, so that $L$ is not conserved, and neutrinos are Majorana particles.

\newcommand{\nbb}{$0\nu\beta\beta$}

Confirmation that $L$ is not conserved would come from the observation of neutrinoless double beta decay (\nbb), the process Nucl $\ra$ Nucl$^\prime + 2e^-$, in which one nucleus decays to another with the emission of two electrons and nothing else. This process, with no leptons in the initial state but two in the final state, is manifestly lepton-number nonconserving, so its discovery would establish that neutrinos are Majorana particles.

\begin{figure}[!hbtp]
\begin{center}
\includegraphics[scale=0.8]{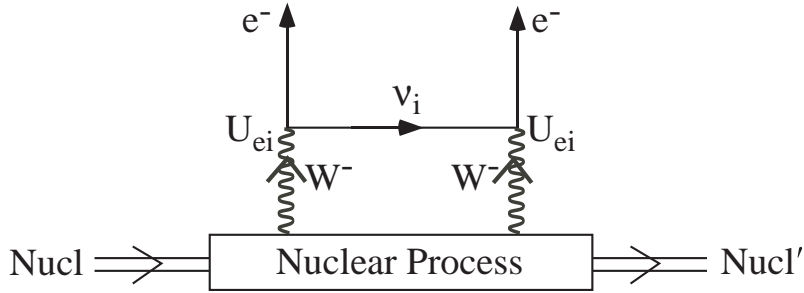}
\caption{Neutrinoless double beta decay.} 
\label{f6} 
\end{center}
\end{figure}
It is expected that the process will be dominated by the diagram shown in Fig.~\ref{f6}. In this diagram, one or another of the neutrino mass eigenstates $\nu_i$ is exchanged between two virtual $W$ bosons to create the outgoing electrons. The \nbb\ amplitude is then a coherent sum over the contributions of the different $\nu_i$.

It is assumed that the $e\nu_i W$ vertices in Fig.~\ref{f6} are Standard Model vertices, involving the SM left-handed weak current. The exchanged $\nu_i$ produced in conjunction with an $e^-$ by this current will have the helicity we normally associate with an antineutrino. That is, it will be dominantly right handed. However, just as the positron emitted in a $\beta$ decay is not 100\% polarized, the virtual $\nu_i$ produced in the process of Fig.~\ref{f6} will not be 100\% polarized either. 
Rather, it will have a small component that is left handed. This component will be of order the mass $m_i$ of $\nu_i$, divided by its energy $E$. It is only this $\mathcal{O}[m_i / E]$ left-handed component that the left-handed current absorbing the exchanged $\nu_i$ can absorb without further suppression. Thus, the contribution of $\nu_i$ exchange to the \nbb\ amplitude is proportional to $m_i$. It is also proportional to $U_{ei}^2$, since, as shown in Fig.~\ref{f6}, there is a factor of $U_{ei}$ at each of the  $e\nu_i W$ vertices. Summing over the contributions of the different $\nu_i$, we conclude that the amplitude for \nbb\ is proportional to the quantity
\beq
m_{\beta\beta} \equiv \left| \sum_i m_i U_{ei}^2 \right| ~~ .
\label{eq52}
\eeq
This quantity is known as the effective Majorana neutrino mass for neutrinoless double beta decay.

That the \nbb\ amplitude growing out of the diagram in Fig.~\ref{f6} is proportional to neutrino mass is no surprise. After all, we have assumed that the $e\nu_i W$ vertices in Fig.~\ref{f6} are SM vertices. But SM vertices conserve $L$. Thus, the $L$-nonconservation required by \nbb\ must be coming from underlying Majorana neutrino mass terms. If the neutrino masses are turned off, the $L$- nonconservation disappears.

It is easy to show that, regardless of what diagrams are actually involved in \nbb, the observation of this decay would still imply the existence of Majorana neutrino mass terms \cite{r20}. Thus, the observation of \nbb\ would teach us that the physics underlying neutrino masses is of a different character than that underlying the masses of the quarks and charged leptons, none of which can have Majorana masses \cite{r21}.

\begin{itemize}
\item Do neutrino interactions violate CP? Is neutrino CP violation the reason we exist?
\end{itemize}

There are several compelling reasons to search for CP violation in neutrino interactions. First, we would like to know whether the leptonic interactions, like the quark interactions, violate CP invariance, or whether CP noninvariance is something peculiar to quarks. Secondly, we would like to know whether leptonic CP violation is responsible for the fact that the universe contains matter (of which we are made) but essentially no antimatter (which, if present, would annihilate us). That is, does leptonic CP violation explain why we exist?

Symmetry and other considerations suggest that, initially, the Big Bang produced equal amounts of matter and antimatter. Thus, the presently-observed preponderance of matter over antimatter must have developed after the initial instants. Now, if one starts with equal amounts of matter and antimatter, and the two behave identically, then one will continue to have equal amounts of the two. Thus, the development of an asymmetry---the present preponderance of matter over antimatter---requires that the two behave differently. That is, it requires a violation of CP invariance.

Experiments with $K$ and $B$ mesons have found CP violation in {\em quark} interactions. However, apart from some unconfirmed anomalies, this quark CP violation is very well described by the Standard Model, and it is known that SM quark CP violation is completely inadequate to explain the observed preponderance of matter over antimatter in the universe. Hence, it is very interesting to ask whether {\em leptonic} CP violation could explain it.

There is a very natural way in which leptonic CP violation could indeed explain the observed matter-antimatter asymmetry. The most popular theory of why neutrinos are so light is the See-Saw Mechanism \cite{r22}. This associates with each light neutrino $\nu$ a very heavy neutrino $N$. The heavier $N$ is, the lighter its see-saw partner, $\nu$, will be (hence the name ``see-saw''). 
The masses of the heavy neutrinos $N$ are thought to be in the range $10^{(9-15)}$ GeV. Thus, these heavy neutrinos cannot be produced in laboratory experiments. However, like everything else, they would have been produced in the hot Big Bang. Now, it is a signature feature of the see-saw model that both the light neutrinos $\nu$ and their heavy see-saw partners $N$ are Majorana particles. Thus, an $N$ can decay both via $N \ra \ell + \dots$ and via $N \ra \bar{\ell} + \dots$, where $\ell$ is a charged lepton. However, if CP is violated in these CP-mirror-image leptonic decays, then
\beq
\Gamma [N \ra \ell + \dots] \neq \Gamma[N \ra \bar{\ell} + \dots] ~~ .
\label{eq53}
\eeq
This rate inequality would have led to an early universe containing unequal numbers of leptons and antileptons. Non-perturbative Standard Model ``sphaleron'' processes would then have converted some of this lepton-antilepton asymmetry into a baryon-antibaryon asymmetry, resulting in the matter-antimatter asymmetric universe that we see today.

The production of the matter-antimatter asymmetry via CP violation in the decays of heavy neutrinos is known as Leptogenesis \cite{r23}. Obviously, we cannot confirm this scenario by repeating the early universe. However, we can lend credibility to it by demonstrating that CP is violated in the interactions of today's light neutrinos $\nu$, which are the see-saw partners of the heavy neutrinos. We can demonstrate this light-neutrino CP violation by showing that the probabilities for the CP-mirror-image oscillations $\nu_\alpha \ra \nu_\beta$ and $\overline{\nu_\alpha} \ra \overline{\nu_\beta}$ are different.

It should be noted that even when, as in the see-saw model, the neutrino mass eigenstates are identical to their antiparticles, $\nu_\alpha \ra \nu_\beta$ and ``$\overline{\nu_\alpha} \ra \overline{\nu_\beta}$'' are still different processes. For example, when we make neutrinos via $\pi^+$ decay, and look for the production of electrons by these neutrinos in a detector, we are studying the oscillation $\nu_\mu \ra \nu_e$. 
However, when we make neutrinos via $\pi^-$ decay, and look for the production of positrons by these neutrinos, we are studying the oscillation conventionally referred to as ``$\overline{\nu_\mu} \ra \overline{\nu_e}$''. It is called $\overline{\nu_\mu} \ra \overline{\nu_e}$ because it would involve antineutrinos if the latter differed from neutrinos. Regardless of notation, the oscillation involving a $\pi^+$ at the source and an $e^-$ at the detector, and its CP-mirror image, the oscillation involving a $\pi^-$ at the source and an $e^+$ at the detector, are clearly different processes. 
If the probabilities for these two processes, which are given by \Eq{15} even when the neutrino mass eigenstates are identical to their antiparticles \cite{r11}, differ, then CP is violated.

The relation between CP violation in (light) neutrino oscillation and in heavy $N$ decay is model dependent. However, if CP is violated in the oscillation of the light neutrinos, then quite likely it is also violated in the decays of the heavy neutrinos $N$ that are the see-saw partners of the light neutrinos \cite{r25}. Then leptogenesis stemming from $N$ decays may well have been the origin of the matter-antimatter asymmetry of the universe.

If $N$ decays did lead to the present proponderance of matter---of which we are made---over antimatter, then we are all descendants of heavy neutrinos.

\section{CONCLUSION}

Wonderful experiments have led to the discovery of neutrino mass. This discovery has raised some very interesting questions, and we would like very much to learn the answers to these questions.

\begin{acknowledgments}
It is a pleasure to thank C. Albright, A. de Gouv\^{e}a, S. Parke, and L. Stodolsky for many useful conversations relevant to the topics of these lectures. Fermilab is operated by Universities Research Association Inc. under Contract No. DE-AC02-76CH03000 with the United States Department of Energy.
\end{acknowledgments}


\end{document}